\documentclass[a4,center,fleqn,twocolumn]{article}
\usepackage[utf8]{inputenc}
\usepackage{color}
\usepackage{graphicx}
\usepackage{subfigure}
\usepackage{amsmath}
\usepackage{url}
\usepackage{cite}
\usepackage{hyperref}
\newcommand{\changed}[1]{{#1}}

\begin{document}

%\subtitle{Structural Bioinformatics}

\title{ProQ3: Improved model quality assessments using Rosetta
  energy terms.}  
\author{Karolis
  Uziela\,$^{1}$, Bj\"orn Wallner\,$^{2}$ and
  Arne Elofsson\,$^{1*}$ \\
  \small $^{1}$ Department of Biochemistry and Biophysics and Science for Life Laboratory, Stockholm University, 171 21 Solna, Sweden and \\
  \small $^{2}$ Department of Physics, Chemistry and Biology (IFM) /Bioinformatics. Link\"oping University, 581 83 Link\"oping, Sweden. \\
  \small $^\ast$To whom correspondence should be addressed.
  }

\maketitle

\begin{abstract}
\textbf{Motivation:} To assess the quality of a protein
  model, i.e. to estimate how close it is to its native structure,
  using no other information than the structure of the model has been
  shown to be useful for structure prediction. The state of the art
  method, ProQ2, is based on a machine learning approach that uses a
  number of features calculated from a protein model. Here, we examine
  if these features can be exchanged with energy terms calculated from
  Rosetta and if a combination of these terms can improve the quality
  assessment. \\
\textbf{Results:} When using the full atom energy function from
Rosetta in ProQRosFA the QA is on par with our previous
state-of-the-art method, ProQ2. The method based on the low-resolution
centroid scoring function, ProQRosCen, performs almost as well and the
combination of all the three methods, ProQ2, ProQRosFA and
ProQCenFA into ProQ3 show superior performance over ProQ2.\\
\textbf{Availability:} ProQ3 is freely available on BitBucket at https://bitbucket.org/ElofssonLab/proq3 \\
\textbf{Contact:} \href{arne@bioinfo.se}{arne@bioinfo.se}\\
%\textbf{Supplementary information:} Supplementary data are available at \textit{Bioinformatics}online.}
\end{abstract}

\section{Introduction}

Protein Model Quality Assessment (MQA) has a long history in protein structure
prediction. Ideally if we could accurately describe the free energy of
a protein this free energy should have a minima at the native
structure, and the free energy could be used to assess the quality of
a protein model. 

Rapid methods to estimate free energies of protein models has been
developed for more than 20
years~\cite{Hendlich2121999,Jones1614539,Luthy1538787}. However the
vast majority of these energy functions were focused on identifying
the native structure among a set of decoys. This means that they do
not necessary show a good correlation with the quality of a protein
model.

In 2003 we set out using a different approach~\cite{Wallner12717029}
with the development of ProQ. Instead of developing a method that
recognised the native structure we developed a method that predicted
the quality of a model. We used a machine learning approach and use a
number of features including agreement with secondary structure,
number and types of atom-atom and residue-residue contacts etc.  One
important reason for the good performance of ProQ was that each type
of contacts, both atom and residue- based ones, was normalised by the
total number of contacts~\cite{Colovos8401235}.

In ProQ the quality was calculated for the entire model.
In 2006 we extended ProQ so that we estimated the quality of each
residue in a protein model, and then we estimated the quality of the
entire model by simply summing up the individual qualities for each
residue~\cite{Wallner16522791}. This method was shown to be rather
successful in CASP5~\cite{Wallner14579343}.

ProQ performed quite well for almost a decade, but some five years ago
one of us developed the successor, ProQ2~\cite{Ray22963006}. The most
important reason for the improved performance of ProQ2 was the use of
profile weights and both local and global features. ProQ2 has since
its introduction remained the superior single model based quality
assessor in CASP~\cite{Kryshtafovych26344049}.

In CASP it has also been shown that another type of quality estimator
clearly are superior to the single model predictors discussed
here. These estimators are based on the consensus approach introduced
by us in CASP5~\cite{Lundstrom11604541,Wallner14579343}. In these
methods the quality of a model, or a residue, is estimated by
comparing how similar it is to models generated by other methods. The
idea is basically that if a protein model is similar to other protein
model it is more likely to be correct. The basis of these methods
is pairwise comparison of a large set of protein models generated for
each target.  Various methods have been developed but the simplest
methods such as 3D-Jury~\cite{Ginalski12761065} and
Pcons~\cite{Wallner16204344} are actually among the best.  The
correlations between estimated and predicted qualities with these
methods are actually higher than the correlation between two different
quality measures~\cite{Wallner17894353}.

A third group of quality assessors also exist, the so called
quasi-single methods~\cite{Pettitt15955780}. These methods take one
single model as input and construct a model ensemble internally based
on the sequence of input model. The quality of the single model is
estimated using consensus with the model ensemble.

It has been known at least since CASP7 that quality assessments with
consensus methods are superior to any other quality assessment
method~\cite{Wallner17894353}. However, it has lately been realised
that these methods have their
limitations~\cite{Kryshtafovych26344049}. Consensus methods are not
better than single model based models at identifying the best possible
model. In particular, when there is one outstanding model, as the
Baker model for taget T0806 in CASP11~\cite{Ovchinnikov26677056}, the
consensus based methods completely fail. Furthermore, a consensus based
quality predictor cannot be used to refine a model or for sampling. Finally, single model-methods can be used in combination with consensus methods to achieve a better performance than either of the approaches~\cite{Kryshtafovych26344049}. Therefore, the development of an improved single model quality assessor is still
needed.

Here, we present the development of ProQ3. ProQ3 is based on the
combination of three independent predictors, ProQ2 and two novel
predictors ProQCenFA and ProQRosFA. Both these two predictors are
trained in a similar way, as ProQ2 but the inputs are different. Here
the inputs come directly from the Rosetta energy functions, either
from the all-atom or the centroid energy function.

\section{Results and Discussion}

\begin{figure*}[ht]
  \centering
  \includegraphics{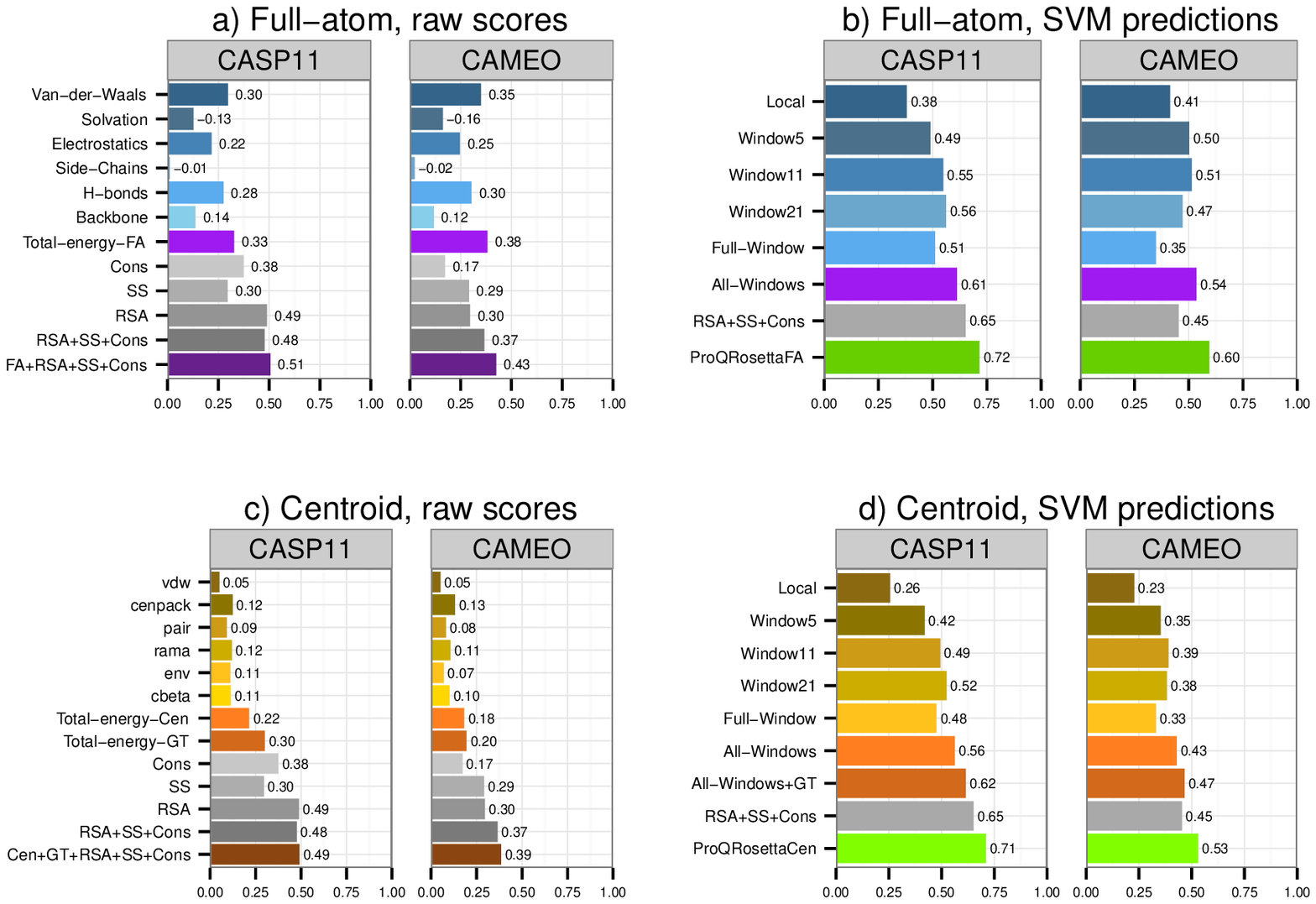}
  \caption{Spearman correlations of ProQRosFA and ProQRosCen input
    features with the target feature (S-score) without SVM training
    (raw scores) and after SVM training (SVM predictions). All the
    correlations are calculated on the local (residue) level. (A)
    Correlations of ProQRosFA input features with S-score. The terms in
    each group are summed up before we calculate the
    correlations. Negative correlations shown with positive bar
    lengths (only Solvation and Side-Chains correlations were
    negative). (B) Correlations of SVM predictions with S-score after
    training SVM on a subset of ProQRosFA input features. Local means
    all the energies are calculated over a single
    residue. Window5-11-21 means that all the energies are averaged
    over a window of 5-11-21 residues. Full-window means that all the
    energies are averaged over the whole model length. All-Windows
    includes all window sizes into the training. RSA, SS and Cons were
    used with the same window sizes as in ProQ2 when training SVM (see
    Methods) (C) Correlations of ProQRosCen input features with
    S-score. Global terms are only presented as a total energy, but not
    individual features. (D) Correlations of SVM predictions with
    S-score after training SVM on a subset of ProQRosCen input
    features. The window definitions are the same as in
    (B). Abbreviations: Total-energy-FA or FA---Total energy of
    full-atom energy terms, Total-energy-Cen or Cen---Total energy of
    centroid local energy terms, Total-energy-GT or GT---Total energy
    of centroid global energy terms, H-bonds---Hydrogen bonds,
    RSA---relative accessible surface area prediction agreement,
    SS---secondary structure prediction agreement, Cons---residue
    conservation. See Table~\ref{table:feature_table} for a more
    detail explanation of the input features. Color scheme: shades of
    blue and purple---Rosetta full atom input features, shades of
    brown, yellow and orange---Rosetta centroid input features, shades
    of grey---sequence-dependent input features, shades of green---final
    predictors }
  \label{fig:FeatureCors}
\end{figure*}

\begin{figure*}[ht]
  \includegraphics{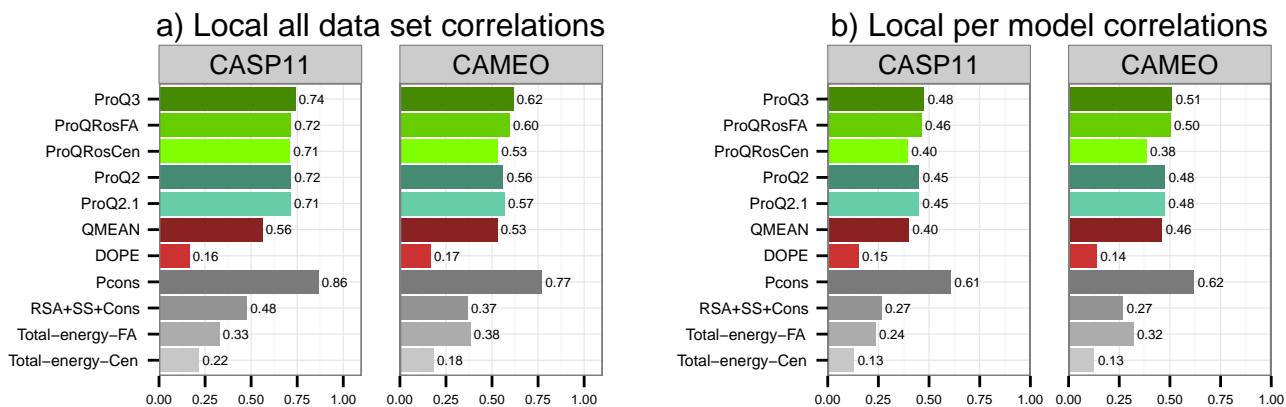}
  \caption{Spearman correlations of selected methods vs. S-score on
    local (residue) level. A) Correlations for the whole data set B)
    Average correlations for each model in the data set. ProQ2.1 is
    ProQ2 retrained on CASP9 data set with 30 models per target (the
    same training data set as ProQ3). RSA+SS+Cons is Relative Surface
    Area accessibility agreement plus Secondary Structure agreement
    plus Conservation without training (same as in
    Figure~\ref{fig:FeatureCors}a~and~\ref{fig:FeatureCors}c). Total-energy-FA
    is the sum of Rosetta full-atom energy terms without training
    (same as in Figure~\ref{fig:FeatureCors}a). Total-energy-Cen is
    the sum of Rosetta centroid energy terms without training (same as
    in Figure~\ref{fig:FeatureCors}c). Color scheme: Shades of
    green---ProQ methods, shades of red---other single-model methods,
    shades of grey---reference methods. }
  \label{fig:local}
\end{figure*}

\begin{figure*}[ht]
  \includegraphics{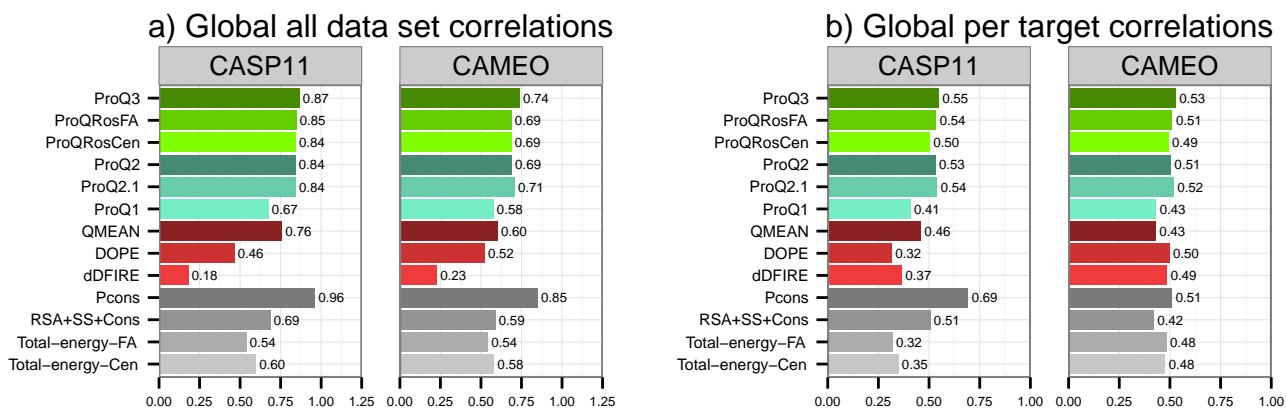}
  \caption{Spearman correlations of selected methods vs. S-score on
    global (protein) level. A) Correlations for the whole data set B)
    Average correlations for each target in the data set. Color scheme
    and method definitions are the same as in
    Figure~\ref{fig:local}. }
  \label{fig:global}
\end{figure*}

\begin{figure*}[ht]
  \includegraphics{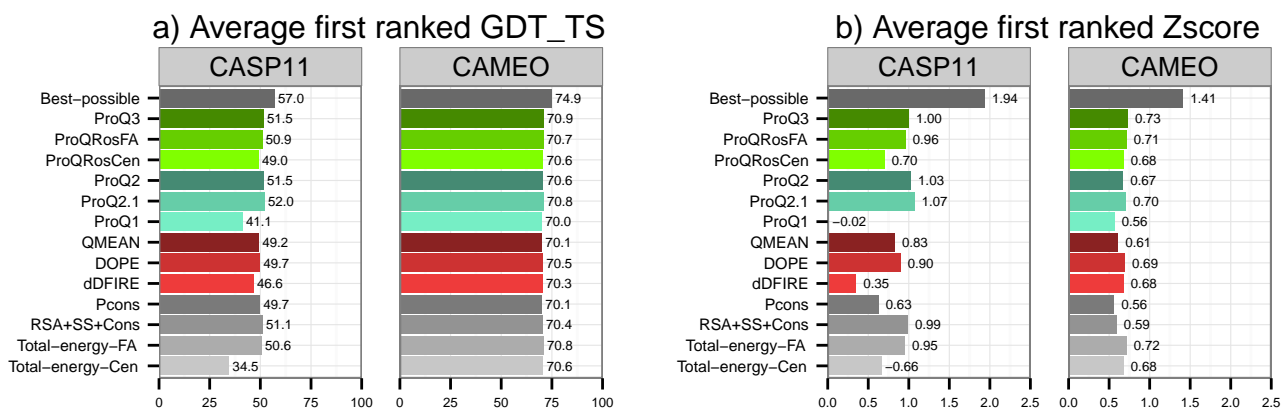}
  \caption{A) Average first ranked GDT\_TS score for each
    method. Average is calculated over all targets in a data set. B)
    Average first ranked Z-score for each method. Z-score is defined as
    $(GDT\_TS_{first\_ranked} - mean(GDT\_TS)) /
    standard\_deviation(GDT\_TS)$
    where mean standard deviation (sd) is calculated for each target.
    Color scheme and method definitions are the same as in
    Figure~\ref{fig:local}.}
  \label{fig:selection}
\end{figure*}

\begin{figure*}[ht]
  \includegraphics{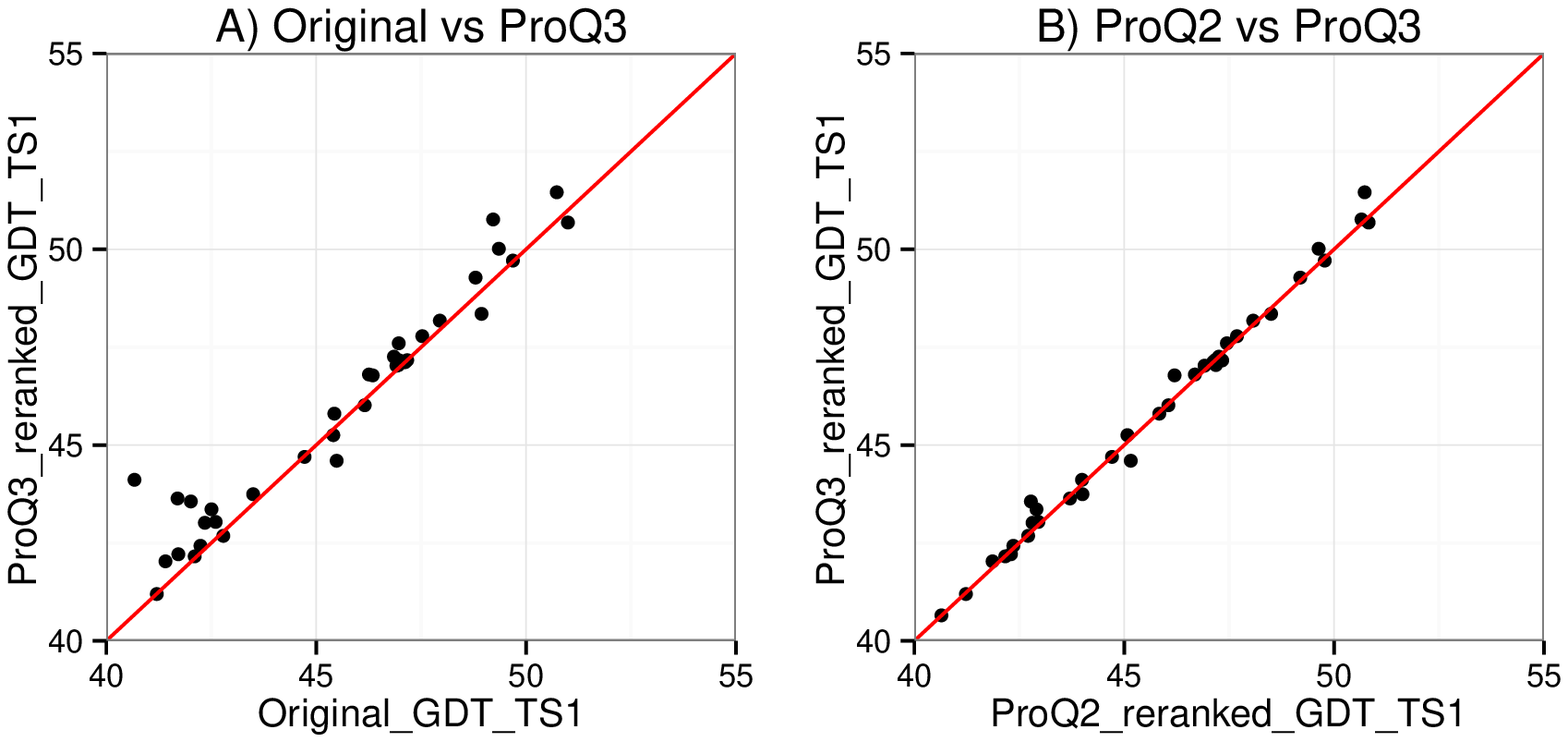}
  \caption{A) Original vs. ProQ3-reranked average GDT\_TS1 scores for each method B) Original vs. ProQ3-reranked average GDT\_TS1 scores for each method. Only methods with average first ranked GDT\_TS greater than 40 are shown (36 out of 44 methods).}
  \label{fig:gdt1_improve}
\end{figure*}

\changed{Our aim of the project was to improve ProQ2 method, which is
  already one of the best, if not the best, single-model methods for
  model quality assessment. ProQ2 is a machine learning based method
  that is based on Support Vector Machines (SVM) that was recently
  implemented as a scoring function in
  Rosetta~\cite{proq2_rosetta}. It has a variety of input features,
  including atom-atom contacts, residue-residue contacts, surface area
  accessibilities, secondary structure and residue conservation.}

\changed{When developing a new predictor we were looking for new input
  features that we could use and decided to use Rosetta energy
  functions. }Rosetta has two types of energy functions: one that uses
an all-atom protein model (``full-atom'' model) and the other one that
uses only a protein backbone (``centroid'' model). In general, the
type of function that uses an all-atom model gives more accurate
results, but the other type can be useful when an all-atom model is
not available. Therefore, we developed two new predictors: one that
uses all-atom model (``ProQRosFA'') and one that uses only the
protein backbone (``ProQRosCen''). In addition, we have
developed a third predictor that combines ProQRosFA,
ProQRosCen and ProQ2 (``ProQ3'').

\changed{The new predictors still use SVMs. They also inherit the
  sequence-dependent features from ProQ2---relative surface area
  accessibility agreement, secondary structure agreement and residue
  conservation. All other input features in ProQRosFA and
  ProQRosCen are Rosetta energy terms. Below we describe the new
  predictors in more detail.}

%The main difference is that atom-atom contacts, residue-residue contacts and absolute values of residue surface area accessibilities and secondary structure are replaced by Rosetta energy terms.

\subsection{ProQRosFA input features}

\begin{table*}[t]
\caption{
{\bf The description of features used in ProQRosFA and ProQRosCen predictors.}}
%\begin{tabular}{|l|l|l|} 
%\begin{tabular}{|p{1.8cm}|p{1.8cm}|p{5cm}|} 
\begin{tabular}{|p{3cm}|p{3cm}|p{6cm}|} 
\hline
\textbf{Feature} & \textbf{Predictor} & \textbf{Description}  \\ 
\hline 
\textbf{Van-der-Waals} & \textbf{ProQRosFA} & A sum of \textit{fa\_atr}, \textit{fa\_rep} and \textit{fa\_intra\_rep} energy terms in Rosetta\\ 
\hline
\textbf{Solvation} & \textbf{ProQRosFA} & \textit{fa\_sol} energy term in Rosetta \\ 
\hline
\textbf{Electrostatics} & \textbf{ProQRosFA} & \textit{fa\_elec} energy term in Rosetta \\ 
\hline
\textbf{Side-chains} & \textbf{ProQRosFA} & \textit{pro\_close}, \textit{dslf\_fa13}, \textit{fa\_dun} and \textit{ref} energy terms in Rosetta \\ 
\hline
\textbf{Hydrogen bond} & \textbf{ProQRosFA} & \textit{hbond\_sr\_bb}, \textit{hbond\_lr\_bb}, \textit{hbond\_bb\_sc} and \textit{hbond\_sc} energy terms in Rosetta \\ 
\hline
\textbf{Backbone} & \textbf{ProQRosFA} & \textit{rama}, \textit{omega} and \textit{p\_aa\_pp } energy terms in Rosetta \\ 
\hline
\textbf{Total-energy-FA} & \textbf{ProQRosFA} & A weighted sum of all energy terms used in ProQRosFA predictor. The weights are the same as in ``Talaris2013''. \\ 
\hline
\textbf{vdw} & \textbf{ProQRosCen} & \textit{vdw} energy term in Rosetta. \\ 
\hline
\textbf{cenpack} & \textbf{ProQRosCen} & \textit{cenpack} energy term in Rosetta. \\ 
\hline
\textbf{pair} & \textbf{ProQRosCen} & \textit{pair} energy term in Rosetta. \\ 
\hline
\textbf{rama} & \textbf{ProQRosCen} & \textit{rama} energy term in Rosetta. \\ 
\hline
\textbf{env} & \textbf{ProQRosCen} & \textit{env} energy term in Rosetta. \\ 
\hline
\textbf{cbeta} & \textbf{ProQRosCen} & \textit{cbeta} energy term in Rosetta. \\ 
\hline
\textbf{Total-energy-Cen} & \textbf{ProQRosCen} & A sum of all \textbf{local} energy terms used in ProQRosCen predictor: \textit{vdw}, \textit{cenpack}, \textit{pair}, \textit{rama}, \textit{env}, \textit{cbeta}. All weights in a sum are equal to one.  \\ 
\hline
\textbf{Total-energy-GT} & \textbf{ProQRosCen} & A sum of all \textbf{global} energy terms used in ProQRosCen predictor: \textit{rg}, \textit{hs\_pair}, \textit{ss\_pair}, \textit{sheet}, \textit{rsigma}, \textit{co}. All weights in a sum are equal to one.  \\ 
\hline
\textbf{RSA} & \textbf{All predictors} & Relative surface area accessibility agreement between the model and the prediction from the sequence.  \\ 
\hline
\textbf{SS} & \textbf{All predictors} & Secondary structure agreement between the model and the prediction from the sequence.  \\ 
\hline
\textbf{Cons} & \textbf{All predictors} & Residue conservation in the sequence \\ 
\hline
\end{tabular}
%\begin{flushleft}
% The description of features used in ProQRosFA and ProQRosCen predictors.
%\end{flushleft}
\label{table:feature_table}
\end{table*}

\begin{table*}[h]
\caption{
{\bf Average GDT\_TS1 for each method before and after reranking }}
%\begin{tabular}{|l|l|l|} 
\begin{tabular}{|p{4cm}|p{2cm}|p{2cm}|p{2cm}|p{2cm}|} 

\hline
\textbf{Method}	& \textbf{Original GDT\_TS1}	& \textbf{ProQ3 GDT\_TS1}	& \textbf{ProQ2 GDT\_TS1}	& \textbf{Optimal GDT\_TS1} \\
\hline
\textbf{QUARK}	& 51.0 &	50.7 &	50.8 &	53.2 \\
\hline
\textbf{Zhang-Server} &	50.7 &	51.5 &	50.7 &	53.4 \\
\hline
\textbf{nns} &	49.7 &	49.7 &	49.8 &	51.7 \\
\hline
\textbf{myprotein-me} &	49.4 &	50.0 &	49.6 &	52.6 \\
\hline
\textbf{BAKER-ROSETTASERVER} &	49.2 &	50.8 &	50.7 &	53.2 \\
\hline

\end{tabular}
\label{table:gdt1_table}
\end{table*}

For ProQRosFA predictor we used ``Talaris 2013'' that is currently
the default score function in Rosetta. This score function consists of
16 terms and the ``Total energy'' term. A SVM model was trained using
all of these energy terms as input features. However, before we
analyse the final performance of ProQRosFA predictor we would like
to show how well the input features were correlated with our target
function \changed{(S-score) without training}. A stronger 
correlation between input feature and the target function is more useful for the final predictor.

Since there are many individual input features (17), rather than
showing the correlation for each separate feature, we will group the
features into 7 groups and show the correlations for each group:
Van-der-Waals, Solvation, Electrostatics, Side-Chains, Hydrogen bond,
Backbone and Total-energy-FA. Note that even though we group features
here for visualising their performance, they were all used separately
when training SVM.

% Here is the list of groups and the features belonging to each group:
%In order to simplify the performance visualization, we grouped the input features into 7 groups:

%\begin{itemize}
%\item \textbf{Van-der-Waals}: fa\_atr, fa\_rep, fa\_intra\_rep 
%\item \textbf{Solvation}: fa\_sol
%\item \textbf{Electrostatics}: fa\_elec
%\item \textbf{Side-chains}: pro\_close, dslf\_fa13, fa\_dun, ref
%\item \textbf{Hydrogen bond}: hbond\_sr\_bb, hbond\_lr\_bb, hbond\_bb\_sc, hbond\_sc
%\item \textbf{Backbone}: rama, omega, p\_aa\_pp 
%\item \textbf{Total-energy-FA}: score \vspace*{1pt}
%\end{itemize}

Figure~\ref{fig:FeatureCors}a shows Spearman correlations against our
target function (S-score) for each of the 7 groups. \changed{On both
  data sets, the correlations for Electrostatics, Van-der-Waals and
  Hydrogen bond and Total-energy-FA groups were higher than Solvation,
  Side-Chains and Backbone.} The Total-energy-FA group includes the
features from all other groups so \changed{it has the highest
  correlation, as expected}. However, the difference in correlations
between Van-der-Waals and Total-energy-FA groups is small.

Relative surface area accessibility agreement (RSA) and secondary
structure agreement (SS) have higher correlation on the CASP11 than on
the CAMEO data set. This maybe due to the fact that models in CAMEO
are higher quality than in CASP11 data set (see
Table~\ref{table:data_set_table}). High quality models usually have
good RSA and SS agreements and, therefore, these features are not as
useful as for lower quality models. Residue conservation (Cons) also
has higher correlation on CASP11 data set and the difference here is
even larger. This is probably because CAMEO targets are more conserved
and are easier to model.

After adding sequence-dependent features (RSA, SS, Cons) to the
Total-energy-FA of Rosetta energies the correlation increases on both
CASP11 and CAMEO data sets even without training. This result is
interesting because it shows that the Rosetta energy potentials could
be improved by sequence-dependent features. However, on CASP11 RSA has
almost as high correlation (0.49) as the combined term FA+RSA+SS+Cons
(0.51), which suggest that RSA alone is a strong feature. In fact, it
is so strong that adding SS and Cons (RSA+SS+Cons) even makes the
correlation slightly smaller than RSA alone.

The correlations can be improved by training SVMs with input features
averaged over a sequence window (see
Figure~\ref{fig:FeatureCors}b). We tried several different window
sizes and observed that on CASP11 the highest correlation (0.56) was
obtained for window of size 21, while on CAMEO the highest correlation
was obtained for window of size 11 (0.51). If all different window
sizes are combined, the correlation increases up to 0.61 on CASP11 and
up to 0.54 on CAMEO data sets. It is interesting that on CASP11 data
set sequence-dependent features alone have a higher correlation than
all the Rosetta features with all window sizes combined (0.65 vs
0.61). This does not hold for CAMEO data set where Rosetta features
have higher correlation than the sequence-dependent features (0.54 vs
0.45). However, when Rosetta features with sequence-dependent features
are combined, the correlation increases on both of the data sets (0.72
and 0.60 on CASP11 and CAMEO respectively).

%In addition to that, the Figure shows the correlations for sequence-dependent features that are calculated outside of Rosetta: RSA---relative accessible surface area prediction agreement, SS---secondary structure prediction agreement, Cons---residue conservation. Finally, it includes ProQRosFA\_NT which is a mean value of Rosetta full-atom energy sum and sequence-dependent features (RSA, SS and Cons). We will use ProQRosFA\_NT and RSA+SS+Cons as reference methods later.

\subsection{ProQRosCen input features}

Centroid scoring functions have an advantage that they can be used
even if the exact position of side chains in the model is not
known. It is also less sensitive to exact atomic positions making it
possible to score models from different methods with a lower risk of high
repulsive score from steric clashes.

The standard centroid scoring function in Rosetta is called
``cen\_std'' and it includes four local centroid energy terms:
\textit{vdw}, \textit{pair}, \textit{env}, \textit{cbeta}. However, it
does not include a couple of other local centroid energy terms
(\textit{cenpack} and \textit{rama}). Therefore, we have defined our
own scoring function that includes all energy terms in ``cen\_std''
plus these two additional ones.

On both CASP11 and CAMEO data sets the two additional energy terms
(\textit{cenpack} and \textit{rama}) have the highest correlation
among all local energy terms, except the Total-energy-Cen term. The
lowest correlation on both data sets is for vdw energy term
(Figure~\ref{fig:FeatureCors}c).

The standard scoring functions in Rosetta ``Talaris2013'' and
``cen\_std'' include only the \textbf{local} energy terms meaning that
these energy terms are defined for each separate residue in the
protein. However, in Rosetta there are also \textbf{global} energy
terms that are only defined for the whole protein model. Most of these
global energy terms are centroid. We have included 6 global centroid
energy terms to our ProQRosCen predictor: \textit{rg},
\textit{hs\_pair}, \textit{ss\_pair}, \textit{sheet}, \textit{rsigma},
\textit{co}. Figure~\ref{fig:FeatureCors}c shows the correlation
between the target function and the Total-energy-Cen of these global
terms (Total-energy-GT). Note that ProQRosCen predictor includes the
same sequence-dependent features as in ProQRosFA. Therefore, the
correlations for sequence-dependent features in
Figure~\ref{fig:FeatureCors}c and Figure~\ref{fig:FeatureCors}a are
the same.

Similarly to what we saw for ProQRosFA, the correlation for the
combined term Cen+GT+RSA+SS+Cons is higher than the correlation for
Total-energy-Cen or Total-energy-GT alone. However, on CASP11 data set
the correlation for RSA is as high as the correlation for the combined
term Cen+GT+RSA+SS+Cons (0.49), which once again confirms that RSA is
a very strong feature.

Just like for ProQRosFA the correlation can be improved by training
SVMs with input features averaged over a sequence window (see
Figure~\ref{fig:FeatureCors}d). On CASP11 data set the highest
correlation is again for window of size 21 (0.52) and on CAMEO data
set the highest correlation is for window of size 11 (0.39). Combining
all window sizes improves the correlation up to 0.56 on CASP11 and up
to 0.43 on CAMEO. Since global energy terms are only defined for the
whole protein, they cannot be averaged over different window
sizes. All-Windows variable in Figure~\ref{fig:FeatureCors}d includes
only local energy terms. When local energy terms averaged over
different window sizes (All-Windows) are combined with global energy
terms (GT), the correlation further increases to 0.62 on CASP11 and
0.47 on CAMEO. After adding sequence-dependent features (RSA+SS+Cons),
the correlation of the final centroid predictor reaches 0.71 on CASP11
and 0.53 on CAMEO.

\subsection{ProQ3} %We should have subsection for ProQ3 for clarity, even though it is short.. -BW

ProQ3 combines the three predictors: ProQ2, ProQRosFA and
ProQRosCen. All the features from three predictors are put together
with the sequence-dependent features that are common to all three
(RSA, SS, Cons) into a new SVM that predicts the score.

\subsection{Benchmark}

In the following sections we will compare ProQ2, ProQRosFA, ProQRosCen
and ProQ3 results with other single-model methods:
QMEAN~\cite{Benkert17932912}, DOPE~\cite{Shen17075131},
DDFIRE~\cite{Yang18260109}, ProQ1~\cite{Wallner12717029}. Only
single-model methods that have publicly available stand-alone versions
were included. Two versions of ProQ2 are included---the original one
and one that was retrained on CASP9 dataset with 30 models per target
(the same training data set as ProQ3). There is also one consensus
method included for a
reference---Pcons~\cite{Wallner16522791}. Finally, sequence-dependent
features and Rosetta total energies are included as a reference.

We will compare the method performance in three categories: local
(residue) level correlations, global (protein) level correlations and
model selection. Two methods (dDFire and ProQ1) provide only the
global level predictions, so they were not included into the local
level evaluation category.

\subsubsection{Local correlations}

All of our new predictors (ProQRosFA, ProQRosCen and ProQ3) are
trained on the local level. In other words, every residue in the
protein model is assigned training feature values and a target value
(S-score). Therefore, it makes sense to look at how well do our
predictions correlate with the target value on the local (residue)
level first.

We have evaluated all methods in two ways: by calculating the
correlation for all data set (Figure~\ref{fig:local}a) and by
calculating the average correlation for each protein model in the data
set (Figure~\ref{fig:local}b). The first evaluation shows how well
methods separate between well-modeled and badly-modeled residues in
general while the second evaluation shows how well methods separate
well-modeled and badly-modeled residues inside a particular model.

ProQ3 outperforms all other single-model methods on both data sets and
in both evaluations. The largest improvement over the original ProQ2
is on CAMEO all data set correlation (0.62 vs. 0.56). ProQRosFA
performs equally or slightly better than the original
ProQ2. ProQRosCen performs slightly worse, but still on par with
QMEAN. Both QMEAN and DOPE perform equally or worse than any ProQ
method with the only exception of QMEAN having a higher per model
correlation than ProQRosCen on CAMEO data set (0.46 vs. 0.38,
Figure~\ref{fig:local}b). The reference consensus method (Pcons)
outperforms all other methods as expected. However, improving
single-model methods is still important for the reasons that were
mentioned in the introduction.  The sequence-dependent features
(RSA+SS+Cons) and Rosetta total energies (Total-energy-FA and
Total-energy-Cen) perform far worse than ProQRosFA and ProQRosCen.
Thus, there is a lot to be gained by using any of any of the methods
developed here.

All differences in local all data set correlations
(Figure~\ref{fig:local}a) are significant with P-values < $10^{-3}$
according to Fisher r-to-z transformation test. All differences in
mean per-model correlations were significant with P-values < $10^{-3}$
according to Wilcoxon signed-rank test.

%When compared to the original ProQ2, the improvement is larger on CAMEO data set than on CASP11 (0.74 vs. 0.72 on CASP11 and 0.62 vs. 0.56 on CAMEO). 

\subsubsection{Global correlations}

Even though ProQRosFA, ProQRosCen and ProQ3 are trained on the local
level, they also provide global predictions of a model quality. The
global predictions are derived from the local predictions, by summing up all
local predictions for a protein model and dividing the sum by the
model length. The target function (S-score) is also local by its
nature, but can be turned to global in exactly the same way.

We have evaluated all methods again in two ways: by calculating the
correlation for all data set (Figure~\ref{fig:global}a) and by
calculating the average correlation for each target in the data set
(Figure~\ref{fig:global}b). The first evaluation shows how well 
a method separates good and bad models in general while the second
evaluation shows how well a method separates good and bad models
for the same target.

ProQ3 again outperforms all other single-model methods on both data
sets and in both evaluations. The largest improvement over the
original ProQ2 is in CAMEO all data set correlation (0.74 vs. 0.69),
Figure~\ref{fig:global}a. Both ProQRosFA and ProQRosCen performance is
close to ProQ2 but better than QMEAN. Like in the local case, the
reference consensus method (Pcons) outperforms all other methods in
most of the evaluations. However, ProQ3 performs better than Pcons on
per target correlations on CAMEO data set (0.53 vs. 0.51),
Figure~\ref{fig:global}b. This shows that consensus methods do not
perform very well on per-target correlations when the number of models
for a target is small (the average number of models per target on
CAMEO data set is only 30, see
Table~\ref{table:data_set_table}). Sequence-dependent features
(RSA+SS+Cons) and raw Rosetta energies (Total-energy-FA and
Total-energy-Cen) again perform worse than ProQRosFA and
ProQRosCen. However, it is interesting to notice that these measures
often perform better than dDFIRE, DOPE and in some cases even QMEAN.

All differences in global all data set correlations
(Figure~\ref{fig:global}a) are significant with P-values < $10^{-3}$
according to Fisher r-to-z transformation test. The number of targets
was too small to get significant differences in per-target correlation
means according to Wilcoxon signed-rank test.

%On the other hand, it is interesting to see that all of these reference methods perform better than DOPE and dDFIRE on all data set correlations. Moreover, these reference features have very good performance in per target correlations. As we already discussed before, on CASP11 data set where the average model quality is bad (see Table~\ref{table:data_set_table}), sequence-dependent features perform well (0.51) and on CAMEO data set where the average model quality is good, raw Rosetta energies perform well (0.48). 

\subsubsection{Model selection}

An important task of model quality assessment program (MQAP) is to be
able to find the best protein model among several possible ones. We
have evaluated MQAP performance in this task by calculating the
average of first ranked GDT\_TS scores (GDT\_TS1) and the average of
first ranked Z-scores for each method (see
Figure~\ref{fig:selection}).

Interestingly, the retrained version of ProQ2 (ProQ2.1) performs
better than ProQ3 both in terms of the average first ranked GDT\_TS
score and Z-scores. Both ProQ2 and ProQ3 outperform all other methods
including Pcons. As shown already at CASP8~\cite{Larsson19544566}, consensus
methods are not performing optimal in model selection and this is one of
the major reasons why we need to develop single-model methods.

After analyzing possible reasons for the better performance of ProQ2
over ProQ3 in model selection, we have found out that ProQ3 is
extremely biased towards selecting Robetta~\cite{Simons9149153}
models. In CASP11, ProQ3 selects Robetta models 65 out of 83 times and
in CAMEO 609 out of 676 times. For comparison, ProQ2 selects Robetta
models 23 times in CASP11 and 444 times in CAMEO.  The reason why
ProQ3 selects Robetta models so often is most likely because Robetta
server models are already optimized using the Rosetta energy function.
Since ProQ3 also uses Rosetta energy terms as input features, it is
not a big surprise that it overestimates quality of Robetta models.
However, this does not harm ProQ3 performance too much, because
Robetta models are often of high quality. Demonstrated by the fact
that 20 targets in CASP11 and 336 targets in CAMEO Robetta models are
the best possible choice. Still, ProQ3 bias towards Robetta models
makes it perform slightly worse than ProQ2 in model selection.
% Could have been even worse if the Robetta models weren't that good ... :-) /BW

\subsubsection{Using ProQ3 to rerank models in structure prediction}

In CASP experiment, structure prediction groups have to submit 5
models for each target and rank them from best to worst. In CASP11
Kryshtafovych~\textit{et al.} drew attention that some of the structure
prediction groups could benefit from using ProQ2 when ranking the
models of their method~\cite{Kryshtafovych26344049}. Similarly to
their analysis, we evaluated how the average GDT\_TS of the first
ranked models would have changed for each group if the structure
prediction groups had been using ProQ2 or ProQ3.

Figure~\ref{fig:gdt1_improve}a shows the average first ranked GDT\_TS
scores for each method before and after reranking them with ProQ3. We
can see that most of the methods (26 out of 36) would have benefited
from using ProQ3 when ranking their models. Only for 8 methods the
score would have gotten worse and for 2 methods it would have stayed
the same.

Figure~\ref{fig:gdt1_improve}a also suggests a ranking of the
structure prediction groups based on the average first ranked
GDT\_TS. If we take all points from right to left we will get the
original ranking of the groups and if we take the points from top to
bottom, we will get the ranking of the groups if all of them were
using ProQ3 to pick the best model out of five. We can see that the
ranking would change for many of the structure prediction groups. In
fact, the ranking would significantly change even for the best
structure prediction groups, as it is shown in
Table~\ref{table:gdt1_table}. If Zhang-Server was using ProQ3, it
would be in the first place instead of second and if
BAKER-ROSETTASERVER was using ProQ3, it would be in the second place
instead of fifth. That clearly shows the benefit of developing good
model quality assessment methods.

%Table~\ref{table:gdt1_table} shows the average GDT_TS1 scores for five best structure prediction groups with and without using ProQ2 or ProQ3 for reranking the models.

Figure~\ref{fig:gdt1_improve}b compares ProQ2 and ProQ3 performance in ranking the models for each structure prediction group. We can see that more than half of the groups (19 out of 36) would have benefited more from using ProQ3 than from using ProQ2. 15 out of 36 groups would have benefited more from using ProQ2 than ProQ3 and for 2 groups there would be no difference.

It is interesting to notice that even BAKER-ROSETTASERVER (Robetta method) group would have benefited more from using ProQ3 than ProQ2. It seems that when models from the same method are evaluated, ProQ3 tendency to overestimate the quality of Rosetta models does not matter. Since all models come from the same method, the bias cancels out and the ranking of the models within the group stays accurate. 

%In order to focus on well performing methods we only show the results for methods that have the average first ranked GDT_TS greater than 40 (36 out of 44 methods). 

%\begin{methods}
\section{Methods}

\subsection{Training and test data sets}

\begin{figure*}[ht]
  \includegraphics{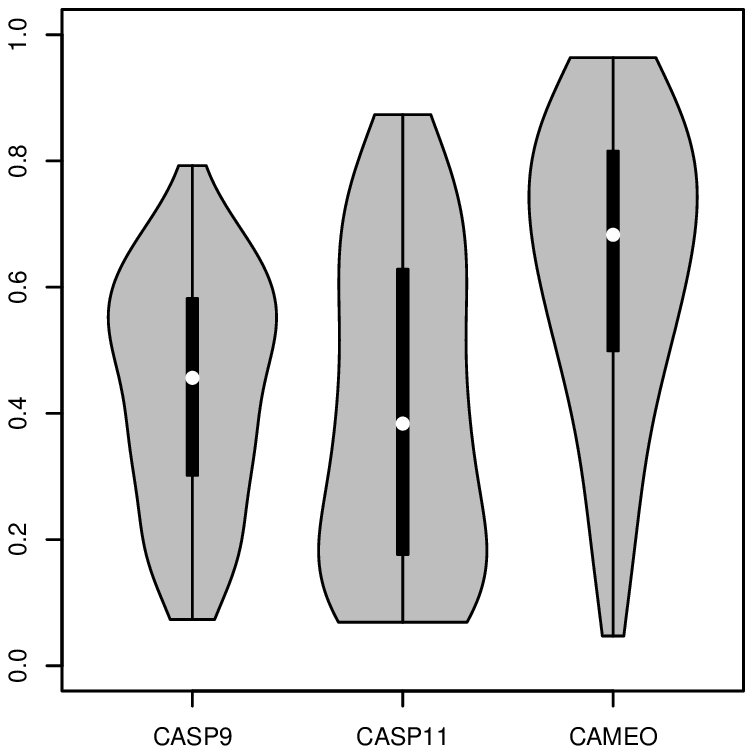}
  \caption{Violin plot of the distribution of mean model quality per target on CASP9, CASP11 and CAMEO data sets.}
  \label{fig:violin}
\end{figure*}

\begin{table*}[h]
\caption{
{\bf Training and test data sets.}}
%\begin{tabular}{|l|l|l|} 
\begin{tabular}{|p{4cm}|p{2cm}|p{2cm}|p{2cm}|p{2cm}|} 
\hline
& \textbf{CASP9} & \textbf{CASP9 random subset} & \textbf{CASP11}  & \textbf{CAMEO} \\ 
\hline 
\textbf{Number of targets} & 	117 & 117 & 83 & 676 \\ 
\hline
\textbf{Total number of models} & 33440 &	3505 &	15334 &	20206 \\
\hline
\textbf{Total number of residues} &	6757370 &	712751 &	3665828 &	5027933 \\
\hline
\textbf{Average number of models per target} &	286 &	30 &	185 &	30 \\
\hline
\textbf{Average number of residues in a model} &	202 &	203 &	239 &	249 \\
\hline
\textbf{Mean model quality} &	0.44 &	0.44 &	0.40 &	0.64 \\
\hline
%\textbf{Standard deviation of model quality} &	0.23 &	0.23 &	0.26 &	0.24 \\
\textbf{\changed{Mean standard deviation of model quality (per target)}} &	0.14  &  0.14  &  0.12  &  0.09 \\
\hline

\end{tabular}

\label{table:data_set_table}
\end{table*}

The original ProQ2 was trained on CASP7 data set with 10 models per target selected at random. We have noticed that the performance slightly increases when ProQ2 is retrained on CASP9 data set with 30 models per target selected at random (see Figure~\ref{fig:local}~and~\ref{fig:global}). Therefore, we have used the latter as the training data set for ProQRosFA, ProQRosCen and ProQ3.

Two data sets were used for testing: CASP11 and CAMEO. Only server models were used in CASP11 data set. All CAMEO models from 1 year time period were used (2014-06-06--2015-05-30). Targets that were shorted than 50 residues were filtered out both from CASP11 and CAMEO data sets. CASP9 data set did not have such short targets.

Table~\ref{table:data_set_table} shows statistics on the data set
sizes and model quality. We can see from the table that CASP9 and
CASP11 data set has more models per target, but CAMEO data set has
more targets. These two facts turn out to compensate each other and
the final number of models is in the same range on all data sets.

Mean model quality on CASP9 and CASP11 data sets is similar (0.44 and 0.40), but on CAMEO data set it is considerably higher (0.64). Mean standard deviation of model quality (calculated per-target) on CAMEO data set is the smallest (0.09). 

Figure~\ref{fig:violin} shows the distribution of mean model quality on CASP9, CASP11 and CAMEO data sets. Mean model quality for most of the targets is clustered around 0.6 on CASP9 data set, around 0.8 on CAMEO data set. CASP11 has more models of bad quality than the other data sets with a small peak around 0.2.

\subsection{Side chain re-sampling and energy minimization}

Protein models can be generated by different methods that employ
different modeling strategies resulting in similar models but vastly
different Rosetta energy terms. For instance, some of models in our
data sets had very large repulsive energy terms (fa\_rep) because of
steric clashes. To account for model generation differences, the
side-chains of all models were rebuild using the backbone-dependent
rotamer library in Rosetta. This was followed by a short backbone
restrained energy minimization protocol using the Rosetta energy
function. This ensured that the Rosetta energy terms for the models
were at their minimium values.

%This improves the MQAP scores for models with poor side chains 

%We have used a similar protocol as Wallner 2014~\cite{} to re-sample protein side chains. 
The side-chains were rebuilt with a backbone-dependent rotamer library implemented in the repack protocol. 10 different decoys were generated for each model and the best one was selected based on ProQ2 score.

Some of the protein models in our data sets had very large Lennard-Jones repulsive energy terms (fa\_rep). To account for this we have run a short energy minimization protocol (-ddg:min\_cst). This has moved protein models from the local maxima of energies without moving the backbones.

\subsection{Implementation}

We have used \textit{per\_residue\_energies} binary in Rosetta (2014
week 5 release) to get per residue energies for local full-atom and
centroid energy terms. \textit{talaris2013.wts} weight file was used
for local full-atom scoring function. For local centroid scoring
function we have defined a custom weight file that included
\textit{vdw}, \textit{cenpack}, \textit{pair}, \textit{rama},
\textit{env}, \textit{cbeta} energy terms with all weights equal to
one.

For global centroid scoring function, Rosetta \textit{score} binary
was used. A custom weight file included \textit{rg},
\textit{hs\_pair}, \textit{ss\_pair}, \textit{sheet}, \textit{rsigma},
\textit{co} energy terms with all weights equal to one.

SVM predictor works best when the input features are either scaled between -1
and 1 or between 0 and 1~\cite{Hsu10apractical}. This is usually achieved by linear scaling of the input
features. However, in order to avoid outliers we decided to use a
sigmoidal function ($1/(1+e^x)$) to scale all of the terms between 0
and 1.

After sigmoidal transformation, all of the local full-atom and
centroid energy terms were averaged using window sizes of 5, 11 and 21
residues. Additionally, the local (single-residue) and the full-window
(averaged over the whole protein) energy terms were added to the
training.

Global centroid energy terms are defined for the whole protein, so
they cannot be averaged using different window sizes. On the other
hand, they depend on the protein size, so they need to be
normalized. \textit{rg} term depends on the protein size L by a factor
of $L^{0.4}$~\cite{Neves1152525} by which it was normalized. After performing a
linear regression on the logarithmic scale we found that \textit{co}
depends on the protein size by $L^{0.72}$ and the other terms by L, so
they were normalized accordingly.

\subsection{Target function}

We have used the same target function as in ProQ2, the S-score. S-score is defined as:
\begin{equation}
S_i = {1 \over {1 + (d_i/d_0)^2}}
\end{equation}
where $d_i$ is the distance for residue i between the native structure and the model in the superposition that maximizes the sum of $S_i$ and $d_0$ is a distance threshold. The distance threshold was set to 3\AA, as in
the original version of ProQ2.

\subsection{Sequence-dependent features}

The sequence-dependent features, RSA, SS and Cons (see
Table~\ref{table:feature_table}) were implemented the same way as in
ProQ2. Sequence profiles were derived using three iterations of
PSI-BLAST v.2.2.26~\cite{Altschul9254694} against Uniref90 (downloaded
2015-10-02)~\cite{Suzek17379688} with a $10^{-3}$ E-value inclusion
threshold. Secondary structure of the protein was calculated from the
model using STRIDE~\cite{Frishman8749853} and predicted from the sequence using
PSIPRED~\cite{Jones10493868}. The agreement between the prediction and the
actual secondary structure in the model was calculated over the window
of 21 residues and over the full-window over the whole protein. Also,
the probability of having a particular secondary structure type in
every single position was calculated. Relative surface area
accessibility was calculated by NACCESS~\cite{citeulike:3431829} and predicted from
the sequence by ACCpro~\cite{Cheng15980571}. The RSA agreement was also
calculated over the window of 21 residues and over the full-window
over the whole protein. The actual secondary structure and relative
surface area was not added to ProQRosFA and ProQRosCen predictors,
only the agreement scores. For residue conservation ``information per
position'' scores were extracted from PSI-BLAST matrix with window
size of 3 residues.

\subsection{SVM training}

A linear SVM model was trained using $SVM^{ligth}$ package V6.02~\cite{joachims2002learning}. All parameters were kept at their default values. 

\subsection{Other tools}

R \textit{zoo} package~\cite{zoo} was used to average values over varying window sizes. \textit{needle} program from EMBOSS package~\cite{Rice10827456} was used to align model and target sequences.

%\end{methods}

%%%%%%%%%%%%%%%%%%%%%%%%%%%%%%%%%%%%%%%%%%%%%%%%%%%%%%%%%%%%%%%%%%%%%%%%%%%%%%%%%%%%%
%
%     please remove the " % " symbol from \centerline{\includegraphics{fig01.eps}}
%     as it may ignore the figures.
%
%%%%%%%%%%%%%%%%%%%%%%%%%%%%%%%%%%%%%%%%%%%%%%%%%%%%%%%%%%%%%%%%%%%%%%%%%%%%%%%%%%%%%%

\section{Conclusion}

Here, we present ProQ3, a novel protein quality prediction
program. ProQ3 is based on a combination of three predictors, ProQ2,
ProQCenFA and ProQRosFA. All these three predictors are trained in a
similar way but the way the inputs are presented for them is
different.  The performance of each individual predictors is similar
and the combination is superior to any of the three individual
predictors.

\section*{Acknowledgements}

We thank Nanjiang Shu for valuable discussions.
%\vspace*{-12pt}

\section*{Funding}

This work was supported by grants from the Swedish  Research Council
(VR-NT 2012-5046 to AE and 2012-5270 to BW). Computational resources
at the National Supercomputing Center 
were provided by SNIC.

%\vspace*{-12pt}

\bibliographystyle{myunsrt}
%\bibliographystyle{achemnat}
%\bibliographystyle{plainnat}
%\bibliographystyle{abbrv}
%\bibliographystyle{bioinformatics}
%
%\bibliographystyle{plain}
%
%\bibliography{Document}

%\begin{thebibliography}{}

\bibliography{proq3_sources}

%\end{thebibliography}
\end{document}